\documentclass[lettersize,journal]{IEEEtran}
\IEEEoverridecommandlockouts
\usepackage{cite}
\usepackage[T1]{fontenc}
\usepackage{graphicx}
\usepackage{amssymb}
\usepackage{amsmath}
\usepackage{amsthm}
\usepackage{subcaption}
\usepackage{microtype}
\usepackage{balance}
\usepackage{xcolor}
\usepackage{algorithm}
\usepackage{algorithmicx}
\usepackage{algpseudocode}
\usepackage[acronym]{glossaries}
\usepackage{url}
\usepackage{float}
\usepackage{booktabs}
\usepackage{hyperref}
\usepackage{graphicx}
\usepackage{balance}

\algrenewcommand\algorithmicindent{0.7em}%

\newacronym{NPRACH}{NPRACH}{narrowband physical random-access channel}
\newacronym{ToA}{ToA}{time of arrival}
\newacronym{CFO}{CFO}{carrier frequency offset}
\newacronym{NBIoT}{NB-IoT}{narrowband internet of things}
\newacronym{5GNR}{5G NR}{5G New Radio}
\newacronym{3GPP}{3GPP}{3rd Generation Partnership Project}
\newacronym{UMi}{UMi}{urban microcell}
\newacronym{RMSE}{RMSE}{root-mean-square error}
\newacronym{NN}{NN}{neural network}
\newacronym{BS}{BS}{base station}
\newacronym{UE}{UE}{user equipment}
\newacronym{SG}{SG}{symbol group}
\newacronym{CP}{CP}{cyclic prefix}
\newacronym{OFDM}{OFDM}{orthogonal frequency division multiplexing}
\newacronym{FFT}{FFT}{fast Fourier transform}
\newacronym{AWGN}{AWGN}{additive white Gaussian noise}
\newacronym{DFT}{DFT}{discrete Fourier transform}
\newacronym{FNR}{FNR}{false negative rate}
\newacronym{FPR}{FPR}{false positive rate}
\newacronym{RG}{RG}{resource grid}
\newacronym{RE}{RE}{resource element}
\newacronym{SNR}{SNR}{signal-to-noise ratio}
\newacronym{1D}{1D}{one-dimensional}
\newacronym{MLP}{MLP}{multilayer perceptron}
\newacronym{BCE}{BCE}{binary cross-entropy}
\newacronym{KL}{KL}{Kullback–Leibler}
\newacronym{SGD}{SGD}{stochastic gradient descent}
\newacronym{ppm}{ppm}{parts-per-million}
\newacronym{ICI}{ICI}{inter-carrier interference}
\newacronym{GNN}{GNN}{graph neural network}
\newacronym{BP}{BP}{belief propagation}
\newacronym{FEC}{FEC}{forward error correction}
\newacronym{LDPC}{LDPC}{low-density parity-check}
\newacronym{HDPC}{HDPC}{high-density parity-check}
\newacronym{SCL}{SCL}{successive cancellation list}
\newacronym{SC}{SC}{successive cancellation}
\newacronym{URLLC}{URLLC}{ultra-reliable low-latency communications}
\newacronym{APP}{APP}{a posterior probability}
\newacronym{MIMO}{MIMO}{multiple-input multiple-output}
\newacronym{CNN}{CNN}{convolutional neural network}
\newacronym{BER}{BER}{bit error rate}
\newacronym{BPSK}{BPSK}{binary phase shift keying}
\newacronym{LLR}{LLR}{log-likelihood ratio}
\newacronym{FN}{FN}{factor node}
\newacronym{VN}{VN}{variable node}
\newacronym{CN}{CN}{check node}
\newacronym{MPNN}{MPNN}{message passing neural network}

\newacronym{AI}{AI}{artificial intelligence}
\newacronym{ML}{ML}{machine learning}
\newacronym{SISO}{SISO}{single input single output}
\newacronym{PRB}{PRB}{physical resource block}
\newacronym{PUSCH}{PUSCH}{physical uplink shared channel}
\newacronym{HARQ}{HARQ}{hybrid automatic repeat request}

\newacronym{MUMIMO}{MU-MIMO}{multi-user multiple-input multiple-output}
\newacronym{BICM}{BICM}{bit-interleaved coded modulation}
\newacronym{QAM}{QAM}{quadrature amplitude modulation}
\newacronym{LMMSE}{LMMSE}{linear minimum mean square error}
\newacronym{MMSE}{MMSE}{minimum mean square error}
\newacronym{CSI}{CSI}{channel-state information}
\newacronym{SIMO}{SIMO}{single-input multiple-output}
\newacronym{CGNN}{CGNN}{convolutional and graph neural network}
\newacronym{BLER}{BLER}{block error rate}
\newacronym{LS}{LS}{least squares}
\newacronym{PE}{PE}{positional encoding}
\newacronym{relu}{ReLU}{rectified linear unit}
\newacronym{RB}{RB}{resource block}
\newacronym{CGGNN}{CGGNN}{convolutional graph neural network}
\newacronym{DMRS}{DMRS}{demodulation reference signal}
\newacronym{IoT}{IoT}{internet of things}
\newacronym{ADAM}{ADAM}{adaptive momentum}
\newacronym{TBLER}{TBLER}{transport block error rate}
\newacronym{MCS}{MCS}{modulation and coding scheme}
\newacronym{TDL}{TDL}{tapped delay line}
\newacronym{CDM}{CDM}{code division multiplexing}
\newacronym{FLOP}{FLOP}{floating point operation}
\newacronym{PHY}{PHY}{physical layer}
\newacronym{MAC}{MAC}{media access control}

\newacronym{ULA}{ULA}{uniform linear array}
\newacronym{NRX}{NRX}{neural receiver}
\newacronym{Var-MCS-NRX}{Var-MCS NRX}{variable-MCS NRX}
\newacronym{UL}{UL}{uplink}
\newacronym{DL}{DL}{downlink}
\newacronym{MSE}{MSE}{mean squared error}
\newacronym{CIR}{CIR}{channel impulse response}
\newacronym{iid}{iid}{independent and identically distributed}

\newacronym{RAN}{RAN}{radio access network}
\newacronym{ORU}{O-RU}{open RAN radio unit}
\newacronym{COTS}{COTS}{commercial-off-the-shelf}
\newacronym{RF}{RF}{radio frequency}
\newacronym{LOS}{LOS}{line-of-sight}
\newacronym{NLOS}{NLOS}{non-line-of-sight}
\newacronym{OTA}{OTA}{over-the-air}
\newacronym{TDD}{TDD}{time-division duplexing}
\newacronym{CAEZ}{CAEZ}{CSI acquisition at ETH Zurich}
\newacronym{ARC-OTA}{ARC-OTA}{Aerial RAN CoLab Over-the-Air}
\newacronym{OAI}{OAI}{OpenAirInterface}
\newacronym{RFFI}{RFFI}{radio frequency fingerprint identification}
\newacronym{FH}{FH}{front haul}
\newacronym{GNSS}{GNSS}{global navigation satellite system}
\newacronym{PTP}{PTP}{precision time protocol}
\newacronym{RNTI}{RNTI}{radio network temporary identifier}
\newacronym{ISM}{ISM}{Industrial, Scientific, and Medical}
\newacronym{UAV}{UAV}{unmanned aerial vehicle}
\newacronym{CDF}{CDF}{cumulative distribution function}
\newacronym{CRC}{CRC}{cyclic redundancy check}
\usepackage{amssymb}
\usepackage{amsfonts}
\usepackage{mathrsfs}
\usepackage{xspace}
\usepackage{bm}
\usepackage{upgreek}

\newcommand{\safemath}[2]{\newcommand{#1}{\ensuremath{#2}\xspace}}

\safemath{\bma}{\mathbf{a}}
\safemath{\bmb}{\mathbf{b}}
\safemath{\bmc}{\mathbf{c}}
\safemath{\bmd}{\mathbf{d}}
\safemath{\bme}{\mathbf{e}}
\safemath{\bmf}{\mathbf{f}}
\safemath{\bmg}{\mathbf{g}}
\safemath{\bmh}{\mathbf{h}}
\safemath{\bmi}{\mathbf{i}}
\safemath{\bmj}{\mathbf{j}}
\safemath{\bmk}{\mathbf{k}}
\safemath{\bml}{\mathbf{l}}
\safemath{\bmm}{\mathbf{m}}
\safemath{\bmn}{\mathbf{n}}
\safemath{\bmo}{\mathbf{o}}
\safemath{\bmp}{\mathbf{p}}
\safemath{\bmq}{\mathbf{q}}
\safemath{\bmr}{\mathbf{r}}
\safemath{\bms}{\mathbf{s}}
\safemath{\bmt}{\mathbf{t}}
\safemath{\bmu}{\mathbf{u}}
\safemath{\bmv}{\mathbf{v}}
\safemath{\bmw}{\mathbf{w}}
\safemath{\bmx}{\mathbf{x}}
\safemath{\bmy}{\mathbf{y}}
\safemath{\bmz}{\mathbf{z}}
\safemath{\bmzero}{\mathbf{0}}
\safemath{\bmone}{\mathbf{1}}
\safemath{\Bell}{\ensuremath{\boldsymbol\ell}}

\bmdefine{\biad}{a}
\bmdefine{\bibd}{b}
\bmdefine{\bicd}{c}
\bmdefine{\bidd}{d}
\bmdefine{\bied}{e}
\bmdefine{\bifd}{f}
\bmdefine{\bigd}{g}
\bmdefine{\bihd}{h}
\bmdefine{\biid}{i}
\bmdefine{\bijd}{j}
\bmdefine{\bikd}{k}
\bmdefine{\bild}{l}
\bmdefine{\bimd}{m}
\bmdefine{\bind}{n}
\bmdefine{\biod}{o}
\bmdefine{\bipd}{p}
\bmdefine{\biqd}{q}
\bmdefine{\bird}{r}
\bmdefine{\bisd}{s}
\bmdefine{\bitd}{t}
\bmdefine{\biud}{u}
\bmdefine{\bivd}{v}
\bmdefine{\biwd}{w}
\bmdefine{\bixd}{x}
\bmdefine{\biyd}{y}
\bmdefine{\bizd}{z}

\bmdefine{\bixid}{\xi}
\bmdefine{\bilambdad}{\lambda}
\bmdefine{\bimud}{\mu}
\bmdefine{\bithetad}{\theta}
\bmdefine{\biphid}{\phi}
\bmdefine{\bideltad}{\delta}

\safemath{\bmia}{\biad}
\safemath{\bmib}{\bibd}
\safemath{\bmic}{\bicd}
\safemath{\bmid}{\bidd}
\safemath{\bmie}{\bied}
\safemath{\bmif}{\bifd}
\safemath{\bmig}{\bigd}
\safemath{\bmih}{\bihd}
\safemath{\bmii}{\biid}
\safemath{\bmij}{\bijd}
\safemath{\bmik}{\bikd}
\safemath{\bmil}{\bild}
\safemath{\bmim}{\bimd}
\safemath{\bmin}{\bind}
\safemath{\bmio}{\biod}
\safemath{\bmip}{\bipd}
\safemath{\bmiq}{\biqd}
\safemath{\bmir}{\bird}
\safemath{\bmis}{\bisd}
\safemath{\bmit}{\bitd}
\safemath{\bmiu}{\biud}
\safemath{\bmiv}{\bivd}
\safemath{\bmiw}{\biwd}
\safemath{\bmix}{\bixd}
\safemath{\bmiy}{\biyd}
\safemath{\bmiz}{\bizd}

\safemath{\bmxi}{\bixid}
\safemath{\bmlambda}{\bilambdad}
\safemath{\bmmu}{\bimud}
\safemath{\bmtheta}{\bithetad}
\safemath{\bmphi}{\biphid}
\safemath{\bmdelta}{\bideltad}

\safemath{\bA}{\mathbf{A}}
\safemath{\bB}{\mathbf{B}}
\safemath{\bC}{\mathbf{C}}
\safemath{\bD}{\mathbf{D}}
\safemath{\bE}{\mathbf{E}}
\safemath{\bF}{\mathbf{F}}
\safemath{\bG}{\mathbf{G}}
\safemath{\bH}{\mathbf{H}}
\safemath{\bI}{\mathbf{I}}
\safemath{\bJ}{\mathbf{J}}
\safemath{\bK}{\mathbf{K}}
\safemath{\bL}{\mathbf{L}}
\safemath{\bM}{\mathbf{M}}
\safemath{\bN}{\mathbf{N}}
\safemath{\bO}{\mathbf{O}}
\safemath{\bP}{\mathbf{P}}
\safemath{\bQ}{\mathbf{Q}}
\safemath{\bR}{\mathbf{R}}
\safemath{\bS}{\mathbf{S}}
\safemath{\bT}{\mathbf{T}}
\safemath{\bU}{\mathbf{U}}
\safemath{\bV}{\mathbf{V}}
\safemath{\bW}{\mathbf{W}}
\safemath{\bX}{\mathbf{X}}
\safemath{\bY}{\mathbf{Y}}
\safemath{\bZ}{\mathbf{Z}}

\safemath{\bZero}{\mathbf{0}}
\safemath{\bOne}{\mathbf{1}}
\safemath{\bDelta}{\mathbf{\Delta}}
\safemath{\bLambda}{\mathbf{\UpLambda}}
\safemath{\bPhi}{\mathbf{\Upphi}}
\safemath{\bSigma}{\mathbf{\Upsigma}}
\safemath{\bOmega}{\mathbf{\Upomega}}
\safemath{\bTheta}{\mathbf{\Uptheta}}

\bmdefine{\biAd}{A}
\bmdefine{\biBd}{B}
\bmdefine{\biCd}{C}
\bmdefine{\biDd}{D}
\bmdefine{\biEd}{E}
\bmdefine{\biFd}{F}
\bmdefine{\biGd}{G}
\bmdefine{\biHd}{H}
\bmdefine{\biId}{I}
\bmdefine{\biJd}{J}
\bmdefine{\biKd}{K}
\bmdefine{\biLd}{L}
\bmdefine{\biMd}{M}
\bmdefine{\biOd}{N}
\bmdefine{\biPd}{O}
\bmdefine{\biQd}{P}
\bmdefine{\biRd}{R}
\bmdefine{\biSd}{S}
\bmdefine{\biTd}{T}
\bmdefine{\biUd}{U}
\bmdefine{\biVd}{V}
\bmdefine{\biWd}{W}
\bmdefine{\biXd}{X}
\bmdefine{\biYd}{Y}
\bmdefine{\biZd}{Z}

\bmdefine{\biDelta}{\Delta}
\bmdefine{\biLambda}{\Lambda}
\bmdefine{\biPhi}{\Phi}
\bmdefine{\biSigma}{\Sigma}
\bmdefine{\biOmega}{\Omega}
\bmdefine{\biTheta}{\Theta}

\safemath{\bimA}{\biAd}
\safemath{\bimB}{\biBd}
\safemath{\bimC}{\biCd}
\safemath{\bimD}{\biDd}
\safemath{\bimE}{\biEd}
\safemath{\bimF}{\biFd}
\safemath{\bimG}{\biGd}
\safemath{\bimH}{\biHd}
\safemath{\bimI}{\biId}
\safemath{\bimJ}{\biJd}
\safemath{\bimK}{\biKd}
\safemath{\bimL}{\biLd}
\safemath{\bimM}{\biMd}
\safemath{\bimN}{\biNd}
\safemath{\bimO}{\biOd}
\safemath{\bimP}{\biPd}
\safemath{\bimQ}{\biQd}
\safemath{\bimR}{\biRd}
\safemath{\bimS}{\biSd}
\safemath{\bimT}{\biTd}
\safemath{\bimU}{\biUd}
\safemath{\bimV}{\biVd}
\safemath{\bimW}{\biWd}
\safemath{\bimX}{\biXd}
\safemath{\bimY}{\biYd}
\safemath{\bimZ}{\biZd}

\safemath{\bimDelta}{\biDelta}
\safemath{\bimLambda}{\biLambda}
\safemath{\bimPhi}{\biPhi}
\safemath{\bimSigma}{\biSigma}
\safemath{\bimOmega}{\biOmega}
\safemath{\bimTheta}{\biTheta}

\safemath{\setA}{\mathcal{A}}
\safemath{\setB}{\mathcal{B}}
\safemath{\setC}{\mathcal{C}}
\safemath{\setD}{\mathcal{D}}
\safemath{\setE}{\mathcal{E}}
\safemath{\setF}{\mathcal{F}}
\safemath{\setG}{\mathcal{G}}
\safemath{\setH}{\mathcal{H}}
\safemath{\setI}{\mathcal{I}}
\safemath{\setJ}{\mathcal{J}}
\safemath{\setK}{\mathcal{K}}
\safemath{\setL}{\mathcal{L}}
\safemath{\setM}{\mathcal{M}}
\safemath{\setN}{\mathcal{N}}
\safemath{\setO}{\mathcal{O}}
\safemath{\setP}{\mathcal{P}}
\safemath{\setQ}{\mathcal{Q}}
\safemath{\setR}{\mathcal{R}}
\safemath{\setS}{\mathcal{S}}
\safemath{\setT}{\mathcal{T}}
\safemath{\setU}{\mathcal{U}}
\safemath{\setV}{\mathcal{V}}
\safemath{\setW}{\mathcal{W}}
\safemath{\setX}{\mathcal{X}}
\safemath{\setY}{\mathcal{Y}}
\safemath{\setZ}{\mathcal{Z}}
\safemath{\emptySet}{\varnothing}

\safemath{\colA}{\mathscr{A}}
\safemath{\colB}{\mathscr{B}}
\safemath{\colC}{\mathscr{C}}
\safemath{\colD}{\mathscr{D}}
\safemath{\colE}{\mathscr{E}}
\safemath{\colF}{\mathscr{F}}
\safemath{\colG}{\mathscr{G}}
\safemath{\colH}{\mathscr{H}}
\safemath{\colI}{\mathscr{I}}
\safemath{\colJ}{\mathscr{J}}
\safemath{\colK}{\mathscr{K}}
\safemath{\colL}{\mathscr{L}}
\safemath{\colM}{\mathscr{M}}
\safemath{\colN}{\mathscr{N}}
\safemath{\colO}{\mathscr{O}}
\safemath{\colP}{\mathscr{P}}
\safemath{\colQ}{\mathscr{Q}}
\safemath{\colR}{\mathscr{R}}
\safemath{\colS}{\mathscr{S}}
\safemath{\colT}{\mathscr{T}}
\safemath{\colU}{\mathscr{U}}
\safemath{\colV}{\mathscr{V}}
\safemath{\colW}{\mathscr{W}}
\safemath{\colX}{\mathscr{X}}
\safemath{\colY}{\mathscr{Y}}
\safemath{\colZ}{\mathscr{Z}}

\safemath{\opA}{\mathbb{A}}
\safemath{\opB}{\mathbb{B}}
\safemath{\opC}{\mathbb{C}}
\safemath{\opD}{\mathbb{D}}
\safemath{\opE}{\mathbb{E}}
\safemath{\opF}{\mathbb{F}}
\safemath{\opG}{\mathbb{G}}
\safemath{\opH}{\mathbb{H}}
\safemath{\opI}{\mathbb{I}}
\safemath{\opJ}{\mathbb{J}}
\safemath{\opK}{\mathbb{K}}
\safemath{\opL}{\mathbb{L}}
\safemath{\opM}{\mathbb{M}}
\safemath{\opN}{\mathbb{N}}
\safemath{\opO}{\mathbb{O}}
\safemath{\opP}{\mathbb{P}}
\safemath{\opQ}{\mathbb{Q}}
\safemath{\opR}{\mathbb{R}}
\safemath{\opS}{\mathbb{S}}
\safemath{\opT}{\mathbb{T}}
\safemath{\opU}{\mathbb{U}}
\safemath{\opV}{\mathbb{V}}
\safemath{\opW}{\mathbb{W}}
\safemath{\opX}{\mathbb{X}}
\safemath{\opY}{\mathbb{Y}}
\safemath{\opZ}{\mathbb{Z}}
\safemath{\opZero}{\mathbb{O}}
\safemath{\identityop}{\opI}

\safemath{\veca}{\bma}
\safemath{\vecb}{\bmb}
\safemath{\vecc}{\bmc}
\safemath{\vecd}{\bmd}
\safemath{\vece}{\bme}
\safemath{\vecf}{\bmf}
\safemath{\vecg}{\bmg}
\safemath{\vech}{\bmh}
\safemath{\veci}{\bmi}
\safemath{\vecj}{\bmj}
\safemath{\veck}{\bmk}
\safemath{\vecl}{\bml}
\safemath{\vecm}{\bmm}
\safemath{\vecn}{\bmn}
\safemath{\veco}{\bmo}
\safemath{\vecp}{\bmp}
\safemath{\vecq}{\bmq}
\safemath{\vecr}{\bmr}
\safemath{\vecs}{\bms}
\safemath{\vect}{\bmt}
\safemath{\vecu}{\bmu}
\safemath{\vecv}{\bmv}
\safemath{\vecw}{\bmw}
\safemath{\vecx}{\bmx}
\safemath{\vecy}{\bmy}
\safemath{\vecz}{\bmz}

\safemath{\veczero}{\bmzero}
\safemath{\vecone}{\bmone}
\safemath{\vecxi}{\bmxi}
\safemath{\veclambda}{\bmlambda}
\safemath{\vecmu}{\bmmu}
\safemath{\vectheta}{\bmtheta}
\safemath{\vecphi}{\bmphi}
\safemath{\vecdelta}{\bmdelta}

\safemath{\matA}{\bA}
\safemath{\matB}{\bB}
\safemath{\matC}{\bC}
\safemath{\matD}{\bD}
\safemath{\matE}{\bE}
\safemath{\matF}{\bF}
\safemath{\matG}{\bG}
\safemath{\matH}{\bH}
\safemath{\matI}{\bI}
\safemath{\matJ}{\bJ}
\safemath{\matK}{\bK}
\safemath{\matL}{\bL}
\safemath{\matM}{\bM}
\safemath{\matN}{\bN}
\safemath{\matO}{\bO}
\safemath{\matP}{\bP}
\safemath{\matQ}{\bQ}
\safemath{\matR}{\bR}
\safemath{\matS}{\bS}
\safemath{\matT}{\bT}
\safemath{\matU}{\bU}
\safemath{\matV}{\bV}
\safemath{\matW}{\bW}
\safemath{\matX}{\bX}
\safemath{\matY}{\bY}
\safemath{\matZ}{\bZ}
\safemath{\matzero}{\bmzero}

\safemath{\matDelta}{\bDelta}
\safemath{\matLambda}{\bLambda}
\safemath{\matPhi}{\bPhi}
\safemath{\matSigma}{\bSigma}
\safemath{\matOmega}{\bOmega}
\safemath{\matTheta}{\bTheta}

\safemath{\matidentity}{\matI}
\safemath{\matone}{\matO}

\safemath{\rnda}{A}
\safemath{\rndb}{B}
\safemath{\rndc}{C}
\safemath{\rndd}{D}
\safemath{\rnde}{E}
\safemath{\rndf}{F}
\safemath{\rndg}{G}
\safemath{\rndh}{H}
\safemath{\rndi}{I}
\safemath{\rndj}{J}
\safemath{\rndk}{K}
\safemath{\rndl}{L}
\safemath{\rndm}{M}
\safemath{\rndn}{N}
\safemath{\rndo}{O}
\safemath{\rndp}{P}
\safemath{\rndq}{Q}
\safemath{\rndr}{R}
\safemath{\rnds}{S}
\safemath{\rndt}{T}
\safemath{\rndu}{U}
\safemath{\rndv}{V}
\safemath{\rndw}{W}
\safemath{\rndx}{X}
\safemath{\rndy}{Y}
\safemath{\rndz}{Z}

\safemath{\rveca}{\bimA}
\safemath{\rvecb}{\bimB}
\safemath{\rvecc}{\bimC}
\safemath{\rvecd}{\bimD}
\safemath{\rvece}{\bimE}
\safemath{\rvecf}{\bimF}
\safemath{\rvecg}{\bimG}
\safemath{\rvech}{\bimH}
\safemath{\rveci}{\bimI}
\safemath{\rvecj}{\bimJ}
\safemath{\rveck}{\bimK}
\safemath{\rvecl}{\bimL}
\safemath{\rvecm}{\bimM}
\safemath{\rvecn}{\bimN}
\safemath{\rveco}{\bomO}
\safemath{\rvecp}{\bimP}
\safemath{\rvecq}{\bimQ}
\safemath{\rvecr}{\bimR}
\safemath{\rvecs}{\bimS}
\safemath{\rvect}{\bimT}
\safemath{\rvecu}{\bimU}
\safemath{\rvecv}{\bimV}
\safemath{\rvecw}{\bimW}
\safemath{\rvecx}{\bimX}
\safemath{\rvecy}{\bimY}
\safemath{\rvecz}{\bimZ}

\safemath{\rvecxi}{\bmxi}
\safemath{\rveclambda}{\bmlambda}
\safemath{\rvecmu}{\bmmu}
\safemath{\rvectheta}{\bmtheta}
\safemath{\rvecphi}{\bmphi}

\safemath{\rmatA}{\bimA}
\safemath{\rmatB}{\bimB}
\safemath{\rmatC}{\bimC}
\safemath{\rmatD}{\bimD}
\safemath{\rmatE}{\bimE}
\safemath{\rmatF}{\bimF}
\safemath{\rmatG}{\bimG}
\safemath{\rmatH}{\bimH}
\safemath{\rmatI}{\bimI}
\safemath{\rmatJ}{\bimJ}
\safemath{\rmatK}{\bimK}
\safemath{\rmatL}{\bimL}
\safemath{\rmatM}{\bimM}
\safemath{\rmatN}{\bimN}
\safemath{\rmatO}{\bimO}
\safemath{\rmatP}{\bimP}
\safemath{\rmatQ}{\bimQ}
\safemath{\rmatR}{\bimR}
\safemath{\rmatS}{\bimS}
\safemath{\rmatT}{\bimT}
\safemath{\rmatU}{\bimU}
\safemath{\rmatV}{\bimV}
\safemath{\rmatW}{\bimW}
\safemath{\rmatX}{\bimX}
\safemath{\rmatY}{\bimY}
\safemath{\rmatZ}{\bimZ}

\safemath{\rmatDelta}{\bimDelta}
\safemath{\rmatLambda}{\bimLambda}
\safemath{\rmatPhi}{\bimPhi}
\safemath{\rmatSigma}{\bimSigma}
\safemath{\rmatOmega}{\bimOmega}
\safemath{\rmatTheta}{\bimTheta}

\usepackage{amssymb}
\usepackage{amsfonts}
\usepackage{mathrsfs}
\usepackage{xspace}
\usepackage{bm}
\usepackage{fancyref}
\usepackage{textcomp}

\usepackage{multirow}
\usepackage{stmaryrd}

\newenvironment{textbmatrix}{	\setlength{\arraycolsep}{2.5pt}%
								\left[\begin{matrix}}{\end{matrix}\right]%
								\raisebox{0.08ex}{\vphantom{M}}}

\def\be{\begin{equation}}
\def\ee{\end{equation}}
\def\een{\nonumber \end{equation}}
\def\mat{\begin{bmatrix}}
\def\emat{\end{bmatrix}}
\def\btm{\begin{textbmatrix}}
\def\etm{\end{textbmatrix}}

\def\ba#1\ea{\begin{align}#1\end{align}}
\def\bas#1\eas{\begin{align*}#1\end{align*}}
\def\bs#1\es{\begin{split}#1\end{split}}
\def\bg#1\eg{\begin{gather}#1\end{gather}}
\def\bml#1\eml{\begin{multline}#1\end{multline}}
\def\bi#1\ei{\begin{itemize}#1\end{itemize}}

\safemath{\dirac}{\delta}					
\safemath{\krond}{\dirac}					

\safemath{\upto}{\uparrow}
\safemath{\downto}{\downarrow}
\safemath{\iu}{j}							
\safemath{\ev}{\lambda}						
\safemath{\hilseqspace}{l^{2}}				
\newcommand{\banachfunspace}[1]{\setL^{#1}}	
\safemath{\hilfunspace}{\banachfunspace{2}}	

\safemath{\SNR}{\textit{SNR}} 				
\safemath{\PAR}{\textit{PAR}} 				
\safemath{\No}{N_0}							
\safemath{\Es}{E_s}							
\safemath{\Eb}{E_b}							
\safemath{\EbNo}{\frac{\Eb}{\No}}
\safemath{\EsNo}{\frac{\Es}{\No}}

\DeclareMathOperator{\CHop}{\ensuremath{\opH}} 
\safemath{\tvir}{\rndh_{\CHop}}				
\safemath{\tvtf}{\rndl_{\CHop}}				
\safemath{\spf}{\rnds_{\CHop}}				
\safemath{\bff}{H_{\CHop}}					

\safemath{\ircf}{r_{h}}						
\safemath{\tftvcf}{r_{s}}					
\safemath{\tfcf}{r_{l}}						
\safemath{\bfcf}{r_{H}}						

\safemath{\tcorr}{c_h}						
\safemath{\scf}{c_{s}}						
\safemath{\tfcorr}{c_{l}}					
\safemath{\fcorr}{c_{H}}						

\safemath{\mi}{I}							
\safemath{\capacity}{C}						

\safemath{\normal}{\mathcal{N}}			
\safemath{\jpg}{\mathcal{CN}}			
\safemath{\mchain}{\leftrightarrow}		

\safemath{\dB}{\,\mathrm{dB}}
\safemath{\dBm}{\,\mathrm{dBm}}
\safemath{\Hz}{\,\mathrm{Hz}}
\safemath{\kHz}{\,\mathrm{kHz}}
\safemath{\MHz}{\,\mathrm{MHz}}
\safemath{\GHz}{\,\mathrm{GHz}}
\safemath{\s}{\,\mathrm{s}}
\safemath{\ms}{\,\mathrm{ms}}
\safemath{\mus}{\,\mathrm{\text{\textmu}s}}
\safemath{\ns}{\,\mathrm{ns}}
\safemath{\ps}{\,\mathrm{ps}}
\safemath{\meter}{\,\mathrm{m}}
\safemath{\mm}{\,\mathrm{mm}}
\safemath{\cm}{\,\mathrm{cm}}
\safemath{\m}{\,\mathrm{m}}
\safemath{\W}{\,\mathrm{W}}
\safemath{\mW}{\, \mathrm{mW}}
\safemath{\J}{\,\mathrm{J}}
\safemath{\K}{\,\mathrm{K}}
\safemath{\bit}{\,\mathrm{bit}}
\safemath{\nat}{\,\mathrm{nat}}

\safemath{\define}{\triangleq}			

\safemath{\equivalent}{\sim}
\safemath{\distas}{\sim}					
\safemath{\sdiff}{\Delta}				

\safemath{\reals}{\mathbb{R}}
\safemath{\positivereals}{\reals_{+}}
\safemath{\integers}{\mathbb{Z}}
\safemath{\posint}{\integers_{+}}
\safemath{\naturals}{\mathbb{N}}
\safemath{\posnaturals}{\naturals_{+}}
\safemath{\complexset}{\mathbb{C}}
\safemath{\rationals}{\mathbb{Q}}

\newcommand*{\fancyrefapplabelprefix}{app}		
\newcommand*{\fancyrefthmlabelprefix}{thm}		
\newcommand*{\fancyreflemlabelprefix}{lem}		
\newcommand*{\fancyrefcorlabelprefix}{cor}		
\newcommand*{\fancyrefdeflabelprefix}{def}		
\newcommand*{\fancyrefproplabelprefix}{prop}		
\newcommand*{\fancyrefexmpllabelprefix}{exmpl}
\newcommand*{\fancyrefalglabelprefix}{alg}		
\newcommand*{\fancyreftbllabelprefix}{tbl}		

\frefformat{vario}{\fancyrefseclabelprefix}{Sec.~#1}
\frefformat{vario}{\fancyrefthmlabelprefix}{Thm.~#1}
\frefformat{vario}{\fancyreftbllabelprefix}{Tbl.~#1}
\frefformat{vario}{\fancyreflemlabelprefix}{Lem.~#1}
\frefformat{vario}{\fancyrefcorlabelprefix}{Corr.~#1}
\frefformat{vario}{\fancyrefdeflabelprefix}{Def.~#1}
\frefformat{vario}{\fancyreffiglabelprefix}{Fig.~#1}
\frefformat{vario}{\fancyrefapplabelprefix}{App.~#1}
\frefformat{vario}{\fancyrefeqlabelprefix}{(#1)}
\frefformat{vario}{\fancyrefproplabelprefix}{Prop.~#1}
\frefformat{vario}{\fancyrefexmpllabelprefix}{Ex.~#1}
\frefformat{vario}{\fancyrefalglabelprefix}{Alg.~#1}

\safemath{\dictab}{[\,\dicta\,\,\dictb\,]}

\safemath{\ysig}{\bmy}
\safemath{\ysighat}{\hat{\ysig}}
\safemath{\ysigdim}{M}
\safemath{\xsig}{\bmx}
\safemath{\xsigdim}{N}
\safemath{\nx}{n_x}
\safemath{\zsig}{\bmz}
\safemath{\zsigdim}{\ysigdim}
\safemath{\rsig}{\bmr}
\safemath{\Adict}{\bA}
\safemath{\Adicttilde}{\widetilde{\Adict}}
\safemath{\Adictdim}{\outputdim\times\xsigdim}
\safemath{\avec}{\bma}
\safemath{\avectilde}{\tilde{\avec}}
\safemath{\Bdict}{\bB}
\safemath{\Bdicttilde}{\widetilde{\Bdict}}
\safemath{\Cdict}{\bC}
\safemath{\cvec}{\bmc}
\safemath{\Ddict}{\bD}
\safemath{\Ddictdim}{\ysigdim\times\xsigdim}
\safemath{\dvec}{\bmd}
\safemath{\Ddicttilde}{\widetilde{\bD}}
\safemath{\Bonb}{\bB}
\safemath{\bvec}{\bmb}
\safemath{\Bonbdim}{\ysigdim\times\ysigdim}
\safemath{\noise}{\bmn}
\safemath{\noisedim}{\ysigim}
\safemath{\err}{\bme}
\safemath{\errdim}{\ysigdim}
\safemath{\errset}{\setE}
\safemath{\nerr}{n_e}
\safemath{\delop}{\bP_\errset}
\safemath{\delopc}{\bP_{{\errset}^c}}

\safemath{\cplxi}{\imath}
\safemath{\cplxj}{\jmath}

\safemath{\dict}{\matD}
\safemath{\inputdim}{N}		
\safemath{\outputdim}{M}		
\safemath{\sparsity}{S}	
\safemath{\inputdimA}{{N_a}}	
\safemath{\inputdimB}{{N_b}}	
\safemath{\elemA}{{n_a}}	
\safemath{\elemB}{{n_b}}	
\safemath{\resA}{\matR_a}	
\safemath{\resB}{\matR_b}	
\safemath{\subD}{\matS} 
\safemath{\subA}{\matS_a} 
\safemath{\subB}{\matS_b} 
\safemath{\dicta}{\matA} 	
\safemath{\dictb}{\matB} 	
\safemath{\hollowS}{H}
\safemath{\hollowA}{H_a}
\safemath{\hollowB}{H_b}
\safemath{\cross}{Z}
\safemath{\coh}{\mu_d}			
\safemath{\coha}{\mu_a}			
\safemath{\cohb}{\mu_b}			
\safemath{\mubs}{\nu}	
\safemath{\cohm}{\mu_m} 
\safemath{\dictset}{\setD}	
\safemath{\dictsetp}{\dictset(\coh,\coha,\cohb)}	
\safemath{\dictsetgen}{\dictset_\text{gen}}
\safemath{\dictsetgenp}{\dictsetgen(\coh)}
\safemath{\dictsetonb}{\dictset_\text{onb}}
\safemath{\dictsetonbp}{\dictsetonb(\coh)}

\safemath{\leftside}{U}
\safemath{\rightsideA}{R_a}
\safemath{\rightsideB}{R_b}

\safemath{\indexS}{\setI_S} 

\safemath{\na}{n_a}			
\safemath{\nb}{n_b}			
\safemath{\coeffa}{p_i}	
\safemath{\coeffb}{q_j}	
\safemath{\seta}{\setP}		
\safemath{\setb}{\setQ}     
\safemath{\setw}{\setW}	
\safemath{\setz}{\setZ}	
\safemath{\cola}{\veca}		
\safemath{\colb}{\vecb}		
\safemath{\cold}{\vecd}		
\safemath{\inputvec}{\vecx} 	
\safemath{\error}{\vece}	
\safemath{\noiseout}{\vecz} 	
\safemath{\inputvecel}{x}
\safemath{\inputveca}{\vecx_a}
\safemath{\inputvecb}{\vecx_b}
\safemath{\outputvec}{\vecy}	
\safemath{\lambdamin}{\lambda_{\mathrm{min}}}

\safemath{\elltwo}{\ell_2}
\safemath{\ellone}{\ell_1}
\safemath{\ellzero}{\ell_0}
\safemath{\ellinf}{\ell_\infty}
\safemath{\ellinftilde}{\ell_{\widetilde\infty}}
\safemath{\licard}{Z(\coh,\coha,\cohb)}
\safemath{\xsol}{\hat{x}}
\safemath{\xbord}{x_b}		
\safemath{\xstat}{x_s}		
\safemath{\xstatLone}{\tilde{x}_s}
\safemath{\order}{\mathcal{O}} 
\safemath{\scales}{\Theta} 
\safemath{\ones}{\mathbf{1}} 
\safemath{\zeroes}{\mathbf{0}} 
\safemath{\thlone}{\kappa(\coh,\cohb)} 
\safemath{\constoneA}{\delta} 
\safemath{\constoneB}{\epsilon} 
\safemath{\nlarge}{L}				   
\safemath{\sumlarge}{S_\nlarge}
\safemath{\maxlarger}{P_\nlarge}	   
\safemath{\Pzero}{\textrm{P0}}	
\safemath{\Pone}{\textrm{P1}}
\safemath{\vecfir}{\vecw}			 
\safemath{\vecsec}{\vecz}
\safemath{\elvecfir}{w}              
\safemath{\elvecsec}{z}				 
\safemath{\nlargefir}{n}
\safemath{\normout}{\gamma}
\safemath{\auxfun}{h}
\safemath{\supp}{\textrm{supp}}

\safemath{\indexa}{\ell}
\safemath{\indexb}{r}
\safemath{\indexc}{i}
\safemath{\indexd}{j}

\safemath{\project}{P}

\safemath{\firstslotset}{\setU_1}  
\safemath{\secondslotset}{\setU_2} 
\safemath{\randomset}{\setS} 

\safemath{\Tran}{\textnormal{T}}
\safemath{\Herm}{\textnormal{H}}

\newcommand*{\fancyreflstlabelprefix}{lst}
\fancyrefaddcaptions{english}{%
  \providecommand*{\freflstname}{Listing}%
}
\frefformat{vario}{\fancyreflstlabelprefix}{%
  \freflstname\fancyrefdefaultspacing#1#3%
}

\begin{document}

\title{Site-Specific Finetuning of Neural Receivers \\ with Real-World 5G NR Measurements}

\author{

    \IEEEauthorblockN{Nuri Berke Baytekin$^\text{1}$, Reinhard Wiesmayr$^\text{1}$, Sebastian Cammerer$^\text{2}$, Chris Dick$^\text{2}$, and Christoph Studer$^\text{1}$}\\[0.2cm]
    \textit{$^\textnormal{1}$ETH Zurich, $^\textnormal{2}$NVIDIA; e-mail: wiesmayr@iis.ee.ethz.ch}\\[0.1cm]
    \thanks{The authors thank Frederik Zumegen for his help during CSI collection.}
    \thanks{We acknowledge NVIDIA for their sponsorship of this research.}
    }

\maketitle
\enlargethispage{2\baselineskip}

\glsresetall
\begin{abstract}
Finetuning wireless receivers to a specific deployment scenario can yield significant error-rate performance improvements without increasing processing complexity. However, site-specific finetuning has so far only been demonstrated on synthetic channel data and lacks real-world benchmarks. In this work, we empirically study site-specific finetuning of neural receivers using real-world 5G NR \gls{PUSCH} data collected with an over-the-air testbed at ETH Zurich across three scenarios: (i) a small laboratory, (ii) a large office floor, and (iii) a high-mobility outdoor environment. Our results confirm substantial error-rate performance improvements from site-specific finetuning, consistent with earlier findings based on synthetic channel data. Moreover, we demonstrate that these improvements generalize across different user-equipment hardware and deployment scenarios.
\end{abstract}

\begin{IEEEkeywords}
5G NR, neural receiver, site-specific finetuning, software-defined testbed, real-world over-the-air measurements.
\end{IEEEkeywords}

\glsresetall

\section{Introduction}

\Gls{ML}-enhanced \gls{PHY} signal processing has the potential to improve next-generation wireless systems \cite{ye2018ofdmdl, oshea2017phydl, doha2025deep}. In particular, \glspl{NRX}---model-based or \gls{NN}-based receivers---can adapt to real-world channel characteristics, hardware impairments, and mobility scenarios of specific deployment sites, and thereby outperform classical algorithms designed for simplistic channel models.
State-of-the-art \glspl{NRX} are pretrained on stochastic channel models \cite{honkala2021deeprx, honkala2021deeprxmimo, cammerer2023neuralreceiver5gnr} and possibly finetuned with site-specific ray-tracing data \cite{wiesmayr2025design, durfee2025sionnart}, where significant improvements have been shown at little finetuning cost and no increase in algorithm complexity \cite{wiesmayr2025design}.
However, training and evaluating \glspl{NRX} purely on synthetic data can lead to overly optimistic results \cite{bock2025wireless}: without real-world benchmarks, there is no guarantee that such receivers work in practice or outperform classical alternatives. To answer these open questions, we finetune and evaluate on real data from an over-the-air communication system.
Although demonstrated here for a specific \gls{NRX}, site-specific finetuning applies to any \gls{ML}-based \gls{PHY} algorithm and we keep our discussion general.

\begin{figure}[tp]
    \centering
    \includegraphics[width=\linewidth]{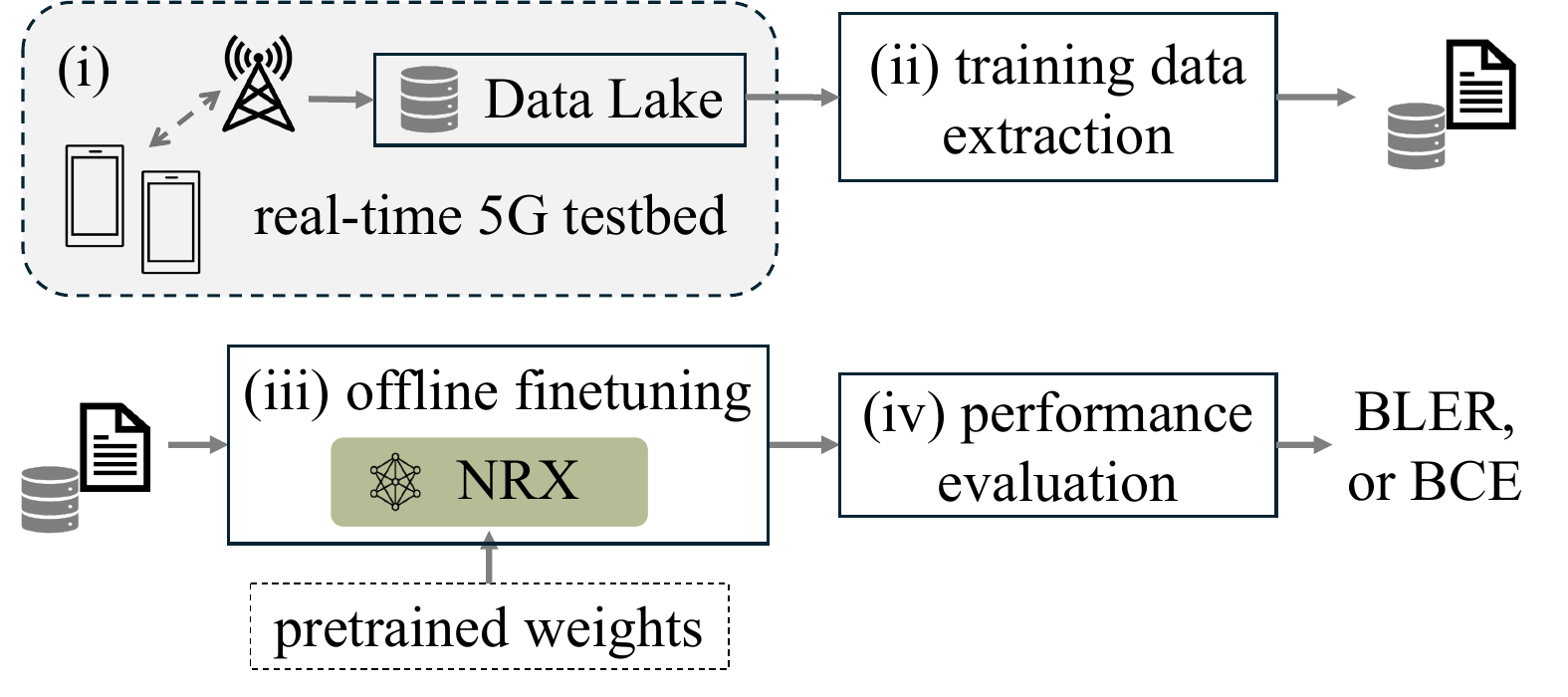}
    \caption{Site-specific finetuning pipeline comprising (i) 5G NR over-the-air measurements in the deployment scenarios shown in \fref{fig:deployment_scenarios}, (ii) training data extraction, (iii) offline site-specific \gls{NRX} finetuning, and (iv) offline real-world block error rate (BLER) performance evaluation on a measured test dataset.}
    \label{fig:schematic}
\end{figure}

\begin{figure*}[tp]
\centering
\begin{subfigure}[t]{0.32\textwidth}
    \centering
    \includegraphics[height=3.75cm]{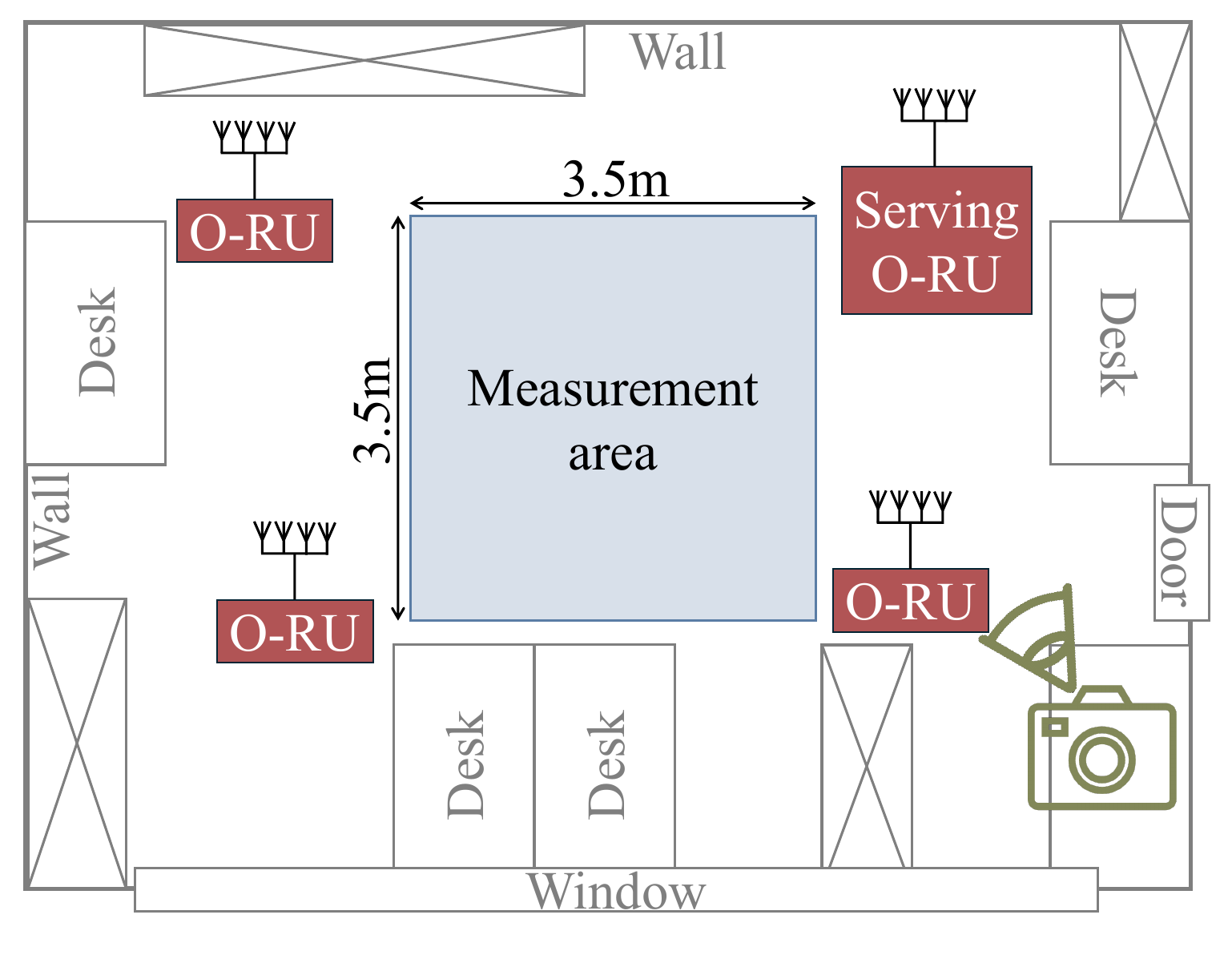}\\[0.2cm]
    \includegraphics[height=3.75cm]{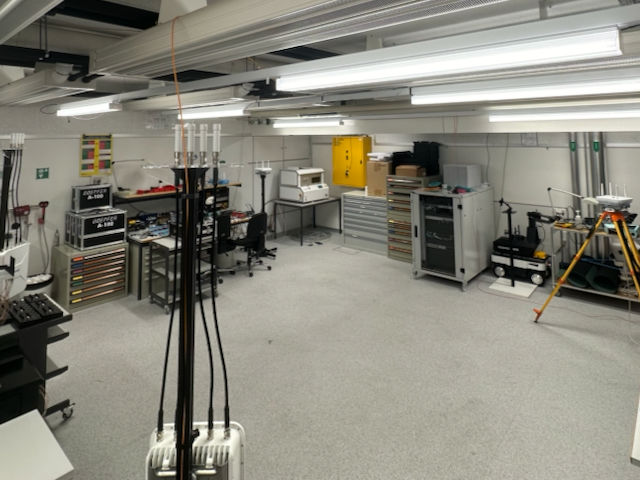}
    \caption{Indoor small laboratory}
    \label{fig:deployment_scenarios_small_lab}
\end{subfigure}
\begin{subfigure}[t]{0.32\textwidth}
    \centering
    \includegraphics[height=3.75cm]{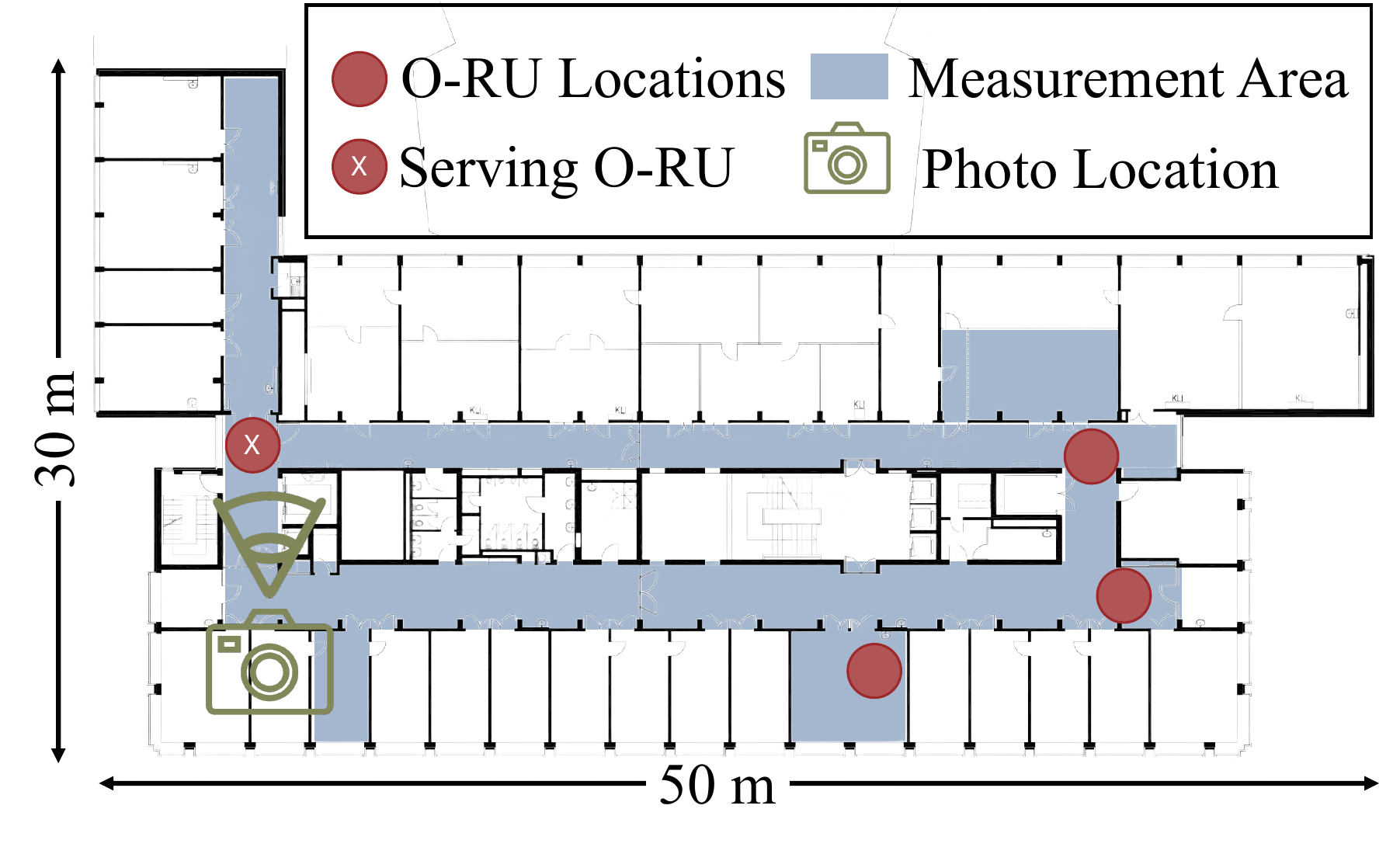}\\[0.2cm]
    \includegraphics[height=3.75cm]{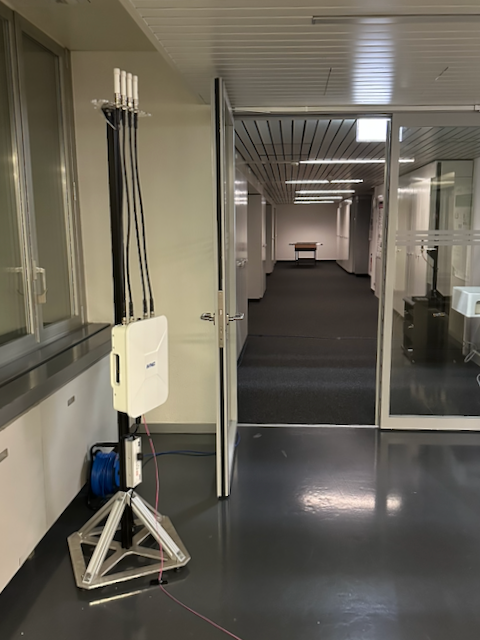}
    \caption{Indoor large office floor}
    \label{fig:deployment_scenarios_institute_floor}
\end{subfigure}
\begin{subfigure}[t]{0.32\textwidth}
    \centering
    \includegraphics[height=3.75cm]{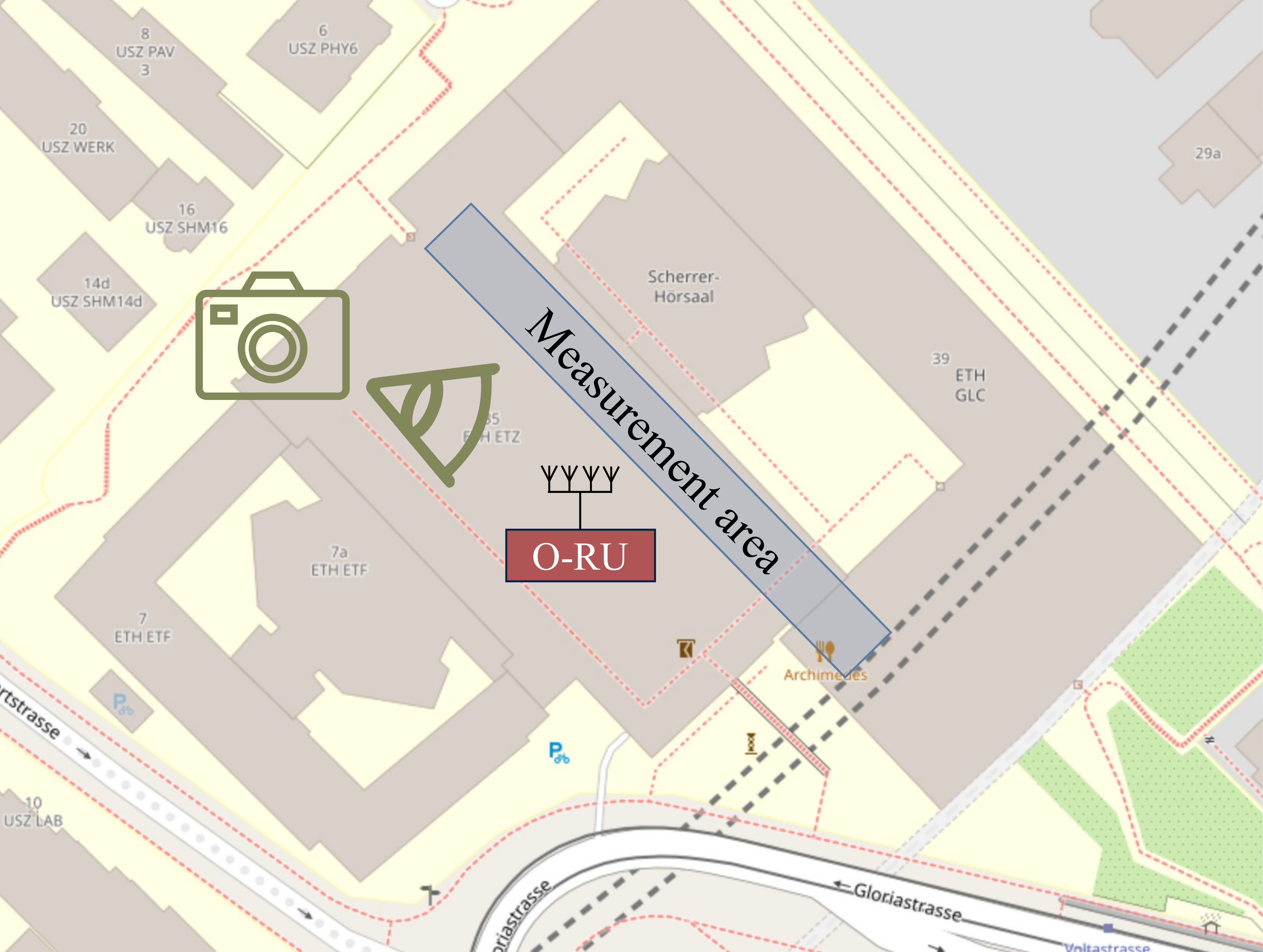}\\[0.2cm]
    \includegraphics[height=3.75cm]{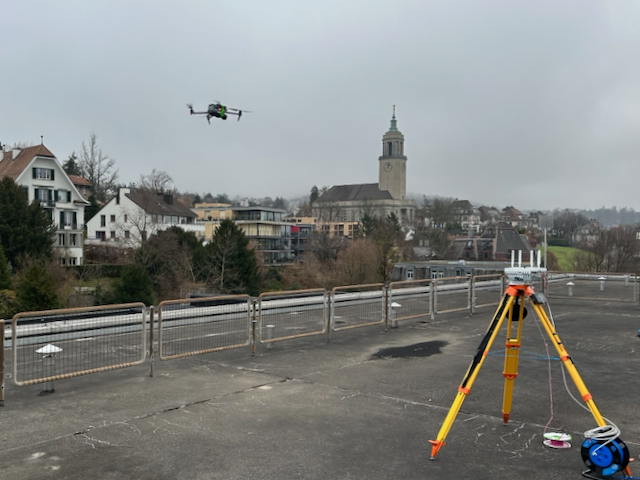}
    \caption{Outdoor high-mobility with UE on UAV}
    \label{fig:deployment_scenarios_outdoor_uav}
\end{subfigure}
\caption{Floor plans (top) and photos (bottom) of the measured deployment scenarios used for site-specific finetuning.}
\label{fig:deployment_scenarios}
\end{figure*}

While site-specific finetuning applies to various \gls{PHY} algorithms---from classical covariance-aided channel estimation~\cite{savaux2017lmmse} to model-based hyperparameter tuning \cite{Wiesmayr2022, abdollahpour2025modelnn}---we focus on \gls{NN}-based receivers \cite{honkala2021deeprxmimo, cammerer2023neuralreceiver5gnr} to investigate the full potential with a fully tunable receiver. Unlike prior work that relied on synthetic ray-tracing data \cite{wiesmayr2025design, durfee2025sionnart, Hoydis2023}, we use measured over-the-air data from the 3GPP-compliant 5G NR testbed at ETH Zurich \cite{wiesmayr2025csi}.
We collect three datasets from \gls{COTS} \glspl{UE} covering (i) a small indoor laboratory, (ii) a large indoor office floor, and (iii) an outdoor high-mobility scenario with a \gls{UE} on an \gls{UAV}. To obtain ground-truth bit labels for failed transmissions, we extend the \gls{NRX} training pipeline from~\cite{wiesmayr2025design} and leverage the 5G NR \gls{HARQ} retransmission mechanism (cf.\ \fref{fig:schematic}).
Using the \gls{NRX} from \cite{cammerer2023neuralreceiver5gnr} in the 5G NR \gls{PUSCH}, we quantify finetuning gains with four experiments:
(i)~\gls{BLER} improvements on measured data,
(ii)~\gls{BLER} improvements for varying \gls{NRX} model complexity,
(iii)~\gls{BLER} improvements for a varying number of finetuning steps, and
(iv)~effective \gls{SNR} improvement with Gaussian noise added to measured baseband data.

\section{Site-specific Deployment Scenarios}

The first step~(i) of our site-specific finetuning pipeline shown in \fref{fig:schematic} is data acquisition. We utilize the 5G NR standard-compliant NVIDIA testbed at ETH~Zurich~\cite{wiesmayr2025csi} for over-the-air measurements with \gls{COTS} \glspl{UE} and \glspl{ORU} in the three deployment scenarios shown in \fref{fig:deployment_scenarios}.

\subsection{Indoor Small Laboratory Scenario}\label{sec:scenario_small_lab}

\fref{fig:deployment_scenarios_small_lab} shows the indoor small laboratory setup with four \glspl{ORU} deployed in the corners. The serving\footnote{The serving \gls{ORU} handles the 5G connection. The other \glspl{ORU} act as passive listeners recording \gls{FH} I/Q samples that can be used for cross-validation. If not stated otherwise, we use data from the serving \gls{ORU}.} \gls{ORU} is in the upper right corner. Data from two \glspl{UE} (Samsung Galaxy S23 and Apple iPhone 14 Pro) are collected by carrying each \gls{UE} along a random trajectory within the measurement area.

\subsection{Indoor Large Office Floor Scenario}

\fref{fig:deployment_scenarios_institute_floor} shows the indoor office floor setup with four \glspl{ORU} and the serving \gls{ORU} in the upper left corner. The same two \glspl{UE} are carried along random trajectories within the measurement area. Compared to the small laboratory, this scenario spans a much larger area and yields a mix of \gls{LOS} and \gls{NLOS} conditions.

\subsection{Outdoor High-Mobility Scenario with \gls{UE} on \gls{UAV}}

\fref{fig:deployment_scenarios_outdoor_uav} shows the outdoor high-mobility setup with one \gls{ORU} on a rooftop at ETH Zurich. A single \gls{UE} (Apple iPhone 16e) is mounted on a DJI Air 3 drone, chosen to comply with the Swiss \gls{UAV} weight limit of 900\,g. The drone flies back and forth at up to 15\,m/s to generate moderate Doppler shifts.

\section{Measurement Setup for Data Extraction}

Next, we describe the training data extraction, i.e., step~(ii) of our site-specific finetuning pipeline shown in \fref{fig:schematic}.

\subsection{System Configuration}\label{sec:sys_config}

Unless stated otherwise, we configure the 5G testbed as in~\cite{wiesmayr2025csi}, i.e., the \gls{TDD} slot pattern 3DSU with 100\,MHz bandwidth (273 \glspl{PRB} at 30\,kHz subcarrier spacing).
To collect meaningful training data for finetuning, we operate the \gls{PUSCH} at a challenging operation point, where we observe transmission errors with the classical \gls{MMSE}-based reference receiver (detailed in \fref{sec:evaluation_pipeline}) that runs in real-time on the 5G testbed. This is important because finetuning on trivial, error-free slots would not yield useful gradient updates for \gls{NRX} training. Therefore, we configure a \gls{PUSCH} \gls{SNR} target of 7\,dB as this was observed to yield approximately 5\% to 10\% \gls{BLER} with \glspl{MCS} using 16-QAM transmission. This operation point was chosen because we apply link adaptation with 10\% \gls{BLER} target and implement the \gls{NRX} architecture for 16-QAM from \cite{cammerer2023neuralreceiver5gnr}.

\subsection{Measurement Procedure}\label{sec:measurement_protocol}
After deploying the 5G testbed, we connect each \gls{UE} and generate application-layer uplink traffic using \texttt{iperf3} while carrying the \gls{UE} along a random trajectory within the measurement area. For campaigns with multiple \glspl{UE}, we measure each \gls{UE} separately.

\subsection{Data Extraction}\label{sec:data_extraction}

During the measurements, NVIDIA Data Lake continuously collects \gls{PUSCH} data \cite{wiesmayr2025csi}: The \gls{FH} table holds raw received I/Q samples from all \glspl{ORU} in the \gls{OFDM} domain, and the FAPI table holds \gls{PHY} and \gls{MAC} layer metadata (scheduling, decoded payload, and \gls{CRC} results) sufficient for offline processing with NVIDIA pyAerial.

\section{Real-World Site-Specific Finetuning} \label{sec:finetuning_pipeline}

The third step~(iii) of our pipeline shown in \fref{fig:schematic} is \gls{NRX} finetuning, which requires training data with ground-truth bit labels from all successful and failed transmissions.

\subsection{Training Data Collection with 5G NR \gls{HARQ} Processing}\label{sec:training_data_collection}
We obtain bit labels of successful transmissions directly from the FAPI table.
For failed transmissions, we leverage the \gls{HARQ} retransmission mechanism \cite{3gpp_nr_rel15}: we track each failed transmission via its \gls{HARQ} process ID and search the FAPI table for the next successful retransmission with the same ID. The corresponding payload bits serve as ground-truth bit labels for the initial failed transmission. Since the FAPI table holds decoded payload bits, we re-encode them with the redundancy version and scrambling configuration of the initial transmission to obtain encoded bit labels for \gls{NRX} finetuning. We monitor the transport block \gls{CRC} to ensure correct decoding.

\subsection{\gls{NRX} Finetuning Training Pipeline}\label{sec:finetuning_training_pipeline}
We leverage the \gls{NRX} finetuning code base from \cite{wiesmayr2025design}. The \gls{FH} data and \gls{LS} channel estimates serve as \gls{NRX} input features, and the \gls{NRX} output \glspl{LLR} are trained with the \gls{BCE} loss against ground-truth bit labels using the Adam optimizer \cite{kingma2015adam} with $10^5$ training batches.
We randomly sample 7500 \gls{PUSCH} slots with 16-QAM and approximately 10\% \gls{BLER} with the default NVIDIA Aerial receiver. As in \cite{cammerer2023neuralreceiver5gnr}, the \gls{NRX} is trained on randomly chosen blocks of four \glspl{PRB} within the 273 \glspl{PRB} per slot, yielding a total of $5.1\cdot10^5$ training samples.

\section{Performance Evaluation}\label{sec:evaluation_pipeline}

The last step (iv) of our pipeline shown in \fref{fig:schematic} is performance evaluation of finetuned and baseline \glspl{NRX}.

\subsection{Performance Baselines}
We compare the following \gls{PHY} receiver algorithms. The ``pretrained \gls{NRX}'' was trained on a randomized 3GPP UMi channel model \cite{38901} using the training configurations specified in~\cite{wiesmayr2025design}\footnote{We pretrained the \gls{NRX} as in \cite{wiesmayr2025design}, except with three \gls{DMRS} symbols (as used by the testbed) instead of two.} with a mixed single- and double-user scenario. The ``finetuned \gls{NRX}'' denotes the \gls{NRX} after site-specific finetuning in the single-user configuration.\footnote{The current 5G testbed only supports single-user transmissions.}
Using the multi-loss training approach from \cite{wiesmayr2025csi} enables us to control the \gls{NN} depth after training by varying the number of \gls{NRX} iterations \cite{cammerer2023neuralreceiver5gnr}. We focus on two configurations: the ``Shallow \gls{NRX}'' (denoted ``Real-Time NRX'' in \cite{wiesmayr2025csi}) with two iterations and the ``Deep \gls{NRX}'' (denoted ``Large NRX'' in \cite{wiesmayr2025csi}) with all eight iterations.
As a baseline, we also evaluate the ``MMSE Reference Rx,'' an advanced receiver implemented in NVIDIA Aerial that applies multi-stage \gls{MMSE} channel estimation with delay estimation and linear \gls{MMSE}-based data detection~\cite{pusch_channel_estimator}.\footnote{We selected the strongest-performing \gls{PUSCH} receiver in NVIDIA Aerial. For direct \gls{NRX} comparisons to other receiver algorithms, see \cite{wiesmayr2025design, cammerer2023neuralreceiver5gnr}.}

\subsection{Evaluation Protocol}

After data detection, all receivers utilize a \gls{LDPC} decoder and operate on an entire slot spanning all 273 \glspl{PRB}. If not stated otherwise, we select the test dataset from slots of the \gls{UE} that was \emph{not} used for training.\footnote{If not described otherwise, we use \gls{PUSCH} slots from UE0 (Samsung S23) for finetuning and slots from UE1 (Apple iPhone 14 Pro) for testing.}
The test dataset comprises 900 slots with matching CRC and 100 slots with CRC failure, sampled to achieve about 10\% \gls{BLER} with the MMSE Reference Rx.

\subsection{Performance Metrics}

\subsubsection{Dataset \gls{BLER}\protect\footnote{The dataset \gls{BLER} is not the ``regular'' \gls{BLER} of the real-time 5G testbed, but the \gls{BLER} of a given receiver on the measured \gls{PUSCH} \gls{FH} signal dataset.}}

This metric characterizes the \gls{BLER} on the measured test dataset after \gls{LDPC} decoding. Unlike a system-level benchmark where link adaptation would adjust the \gls{MCS}, this dataset \gls{BLER} metric enables a direct and fair receiver comparison on the same pre-recorded data.

\subsubsection{\gls{SNR} Improvements with Additive Gaussian Noise}

We quantify \gls{SNR} improvements by adding white Gaussian noise to the received baseband signals. The received signal follows $y=s + n$ with signal component $s\in\complexset$ (received signal power~$E_s$, including pathloss) and noise $n\in\complexset$ (variance $N_0$), recorded at $\gamma=\frac{E_s}{N_0}\approx 7\,\text{dB}$.
We add complex Gaussian noise $z \sim \jpg(0,N_z)$, i.e., $\tilde y = y + z$, with $N_z=\alpha(E_s + N_0)$ and variable $\alpha\geq0$. The \emph{effective \gls{SNR}} is then given by
\begin{equation}
    \text{SNR}_{\text{eff}}
    =\frac{E_s}{N_0+\alpha(E_s + N_0)}
    = \frac{\gamma}{1+\alpha(\gamma+1)}.
\end{equation}
Sweeping $\alpha\geq0$ yields dataset \gls{BLER} vs.\ effective \gls{SNR} curves.

\begin{figure}[t]
    \centering

    \begin{subfigure}{\linewidth}
        \centering
        \includegraphics[width=\linewidth]{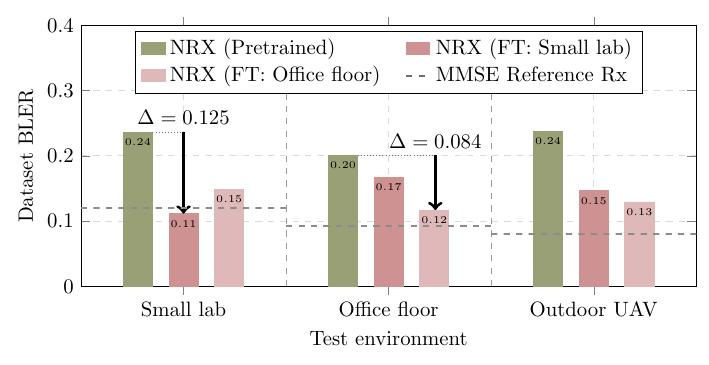}
        \caption{Shallow NRX (2 iter.)}
        \label{fig:results-rt}
    \end{subfigure}

    \vspace{0.8em}

    \begin{subfigure}{\linewidth}
        \centering
        \includegraphics[width=\linewidth]{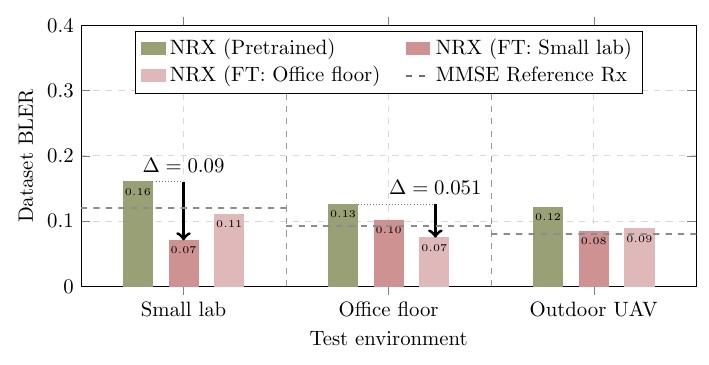}
        \caption{Deep NRX (8 iter.)}
        \label{fig:results-large}
    \end{subfigure}

    \caption{\gls{BLER} performance improvements across scenarios.}
    \label{fig:results-bar}
\end{figure}

\section{Results}\label{sec:results}

We confirm the observations from \cite{wiesmayr2025design} with real-world data. Site-specific finetuning closes the \gls{BLER} gap between the finetuned Shallow \gls{NRX} (2 iter.) and the pretrained Deep \gls{NRX} (8 iter.): The finetuned Shallow \gls{NRX} matches the \gls{MMSE} Reference Rx, while the finetuned Deep \gls{NRX} outperforms it. Furthermore, these performance improvements carry over to different \gls{ORU} deployment scenarios and \glspl{UE}.

\subsection{Dataset \gls{BLER} Improvement}

\fref{fig:results-bar} compares the \gls{BLER} of pretrained and finetuned \glspl{NRX} across all three scenarios for the Shallow \gls{NRX} (2 iter., \fref{fig:results-rt}) and the Deep \gls{NRX} (8 iter., \fref{fig:results-large}). Both subfigures show the \gls{NRX} performance with pretrained weights and finetuned weights for the small laboratory (``FT: Small lab'') and the office floor (``FT: Office floor'').
The finetuned Shallow \gls{NRX} (2 iter.) outperforms the pretrained Deep \gls{NRX} in both indoor environments and even outperforms the advanced MMSE Reference Rx in the small laboratory. The finetuned Deep \gls{NRX} (8 iter.) outperforms the \gls{MMSE} Reference Rx in all scenarios. The \gls{BLER} improvements from finetuning on both indoor scenarios carry over to the other indoor environment with a different \gls{ORU} and to the outdoor high-mobility scenario.

\subsection{Dataset \gls{BLER} Improvement for Varying \gls{NRX} Depth}

\begin{figure}[t]
    \centering
    \includegraphics[width=\linewidth]{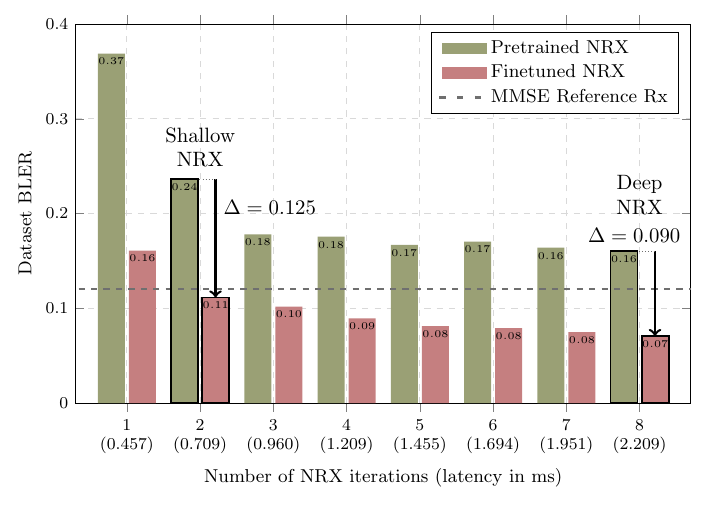}
    \caption{\gls{BLER} performance improvements vs. varying \gls{NRX} complexity for varying number of \gls{NRX} iterations and inference latency.}
    \label{fig:latency}
\end{figure}

\fref{fig:latency} compares the \gls{BLER} of the pretrained and finetuned \gls{NRX} on the indoor small laboratory dataset for a varying number of \gls{NRX} iterations\footnote{We control the \gls{NRX} model complexity by varying the \gls{NN} depth after training, as described in \cite[Sec.~IV~B]{wiesmayr2025csi}.}, shown together with the measured inference latency on an NVIDIA GH200 system. The dataset \gls{BLER} performance of the pretrained \gls{NRX} saturates after two iterations, while the finetuned \gls{NRX} improves up to eight iterations. Finetuning reduces the dataset \gls{BLER} by about one half for all iterations and outperforms the \gls{MMSE} Reference Rx already at two iterations (Shallow \gls{NRX}).

\subsection{Dataset \gls{BLER} Improvements under Varying Number of Finetuning Iterations}

\begin{figure}
    \centering
    \includegraphics[width=\linewidth]{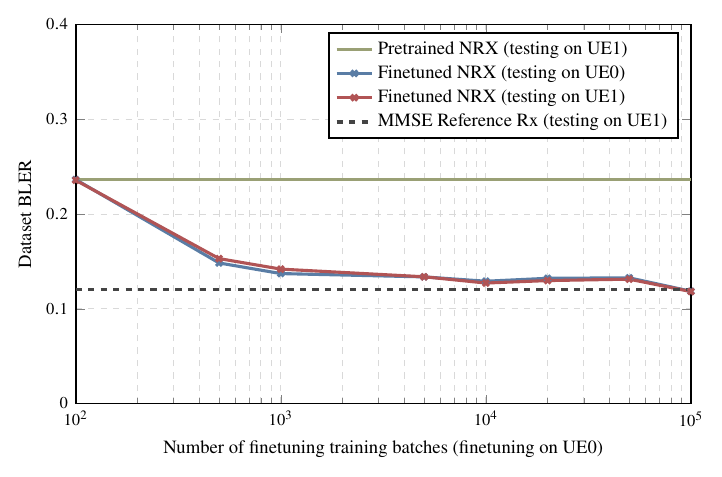}
    \caption{BLER performance vs. number of finetuning training iterations for Shallow \gls{NRX} (2 iter.) finetuned on UE0 (Samsung Galaxy S23).}
    \label{fig:training-iter}
\end{figure}

\fref{fig:training-iter} shows the dataset \gls{BLER} of the finetuned Shallow \gls{NRX} for a varying number of finetuning training batches in the indoor small laboratory. The \gls{NRX} is trained on UE0 (Samsung Galaxy S23) but evaluated on both UE0 and UE1 (Apple iPhone 14 Pro). The similar \gls{BLER} on both \glspl{UE} suggests that the \gls{NRX} does not overfit to a particular device type or 5G modem.\footnote{The Apple iPhone 14 Pro and the Samsung Galaxy S23 have different antenna geometries and 5G modem chipsets (Qualcomm X65 vs. X70).} No improvement is observed in the first $10^2$ batches; significant improvements appear after $10^3$ batches with diminishing returns, and the \gls{MMSE} Reference Rx performance is reached at $10^5$ batches.

\subsection{Effective SNR Improvements}

\begin{figure}[t]
    \centering
    \includegraphics[width=\linewidth]{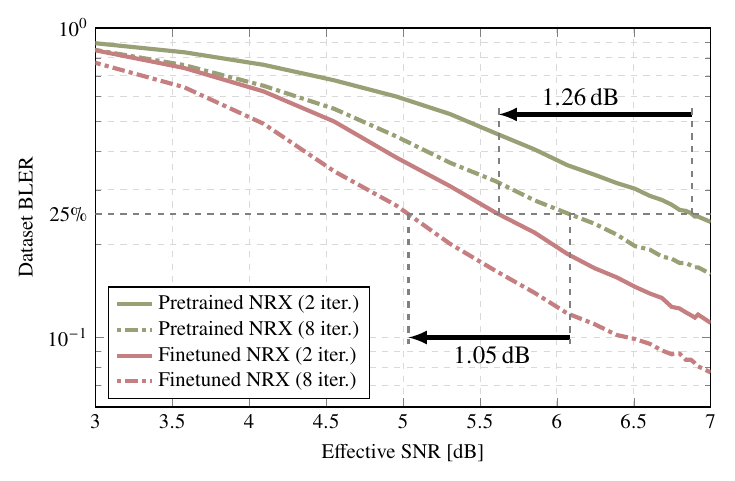}
    \caption{\gls{BLER} performance vs. effective \gls{SNR} with white Gaussian noise added to the indoor small laboratory dataset. \gls{SNR} improvements compared for the minimum \gls{SNR} required to achieve a dataset $\text{BLER}=25\%$.}
    \label{fig:bler-snr}
\end{figure}

\fref{fig:bler-snr} shows the effective \gls{SNR} improvements for the indoor small laboratory. Site-specific finetuning yields an \gls{SNR} gain of 1.26\,dB with the Shallow \gls{NRX} (2 iter.) and 1.05\,dB with the Deep \gls{NRX} (8 iter.). The finetuned Shallow \gls{NRX} even outperforms the pretrained Deep \gls{NRX} at less than $1/3$ of the inference latency (0.7\,ms vs. 2.2\,ms on an NVIDIA GH200).

\section{Conclusions}

We have demonstrated---for the first time---the real-world benefits of site-specific finetuning of \gls{ML}-based \gls{PHY} algorithms across three deployment scenarios. To enable finetuning on real-world data, we proposed a \gls{HARQ}-based method to extract ground-truth bit labels for failed transmissions from over-the-air measurements. Finetuning the \gls{NRX} \cite{cammerer2023neuralreceiver5gnr} reduces the measured dataset \gls{BLER} by more than one half compared to the pretrained counterpart; these improvements carry over to different deployment scenarios, radio units, and user equipment types. Our results with real-world data confirm prior work that used synthetic ray-tracing-based channel data~\cite{wiesmayr2025design}. For example, a finetuned Shallow \gls{NRX} outperforms a pretrained Deep \gls{NRX} at less than $1/3$ the inference latency. Site-specific finetuning yields significant effective \gls{SNR} gains, and meaningful improvements emerge after only a few training batches, highlighting the practicality of the approach. An extension to multi-user scenarios, other modulation orders, model-based receivers \cite{Wiesmayr2022, abdollahpour2025modelnn}, or other low-complexity \gls{NRX} methods \cite{fesl2026learningsic, honkala2026eqdeeprx} is left for future work.

\bibliographystyle{IEEEtran}

\balance

\end{document}